\documentclass[%
 reprint,
 amsmath,amssymb,
prl,
]{revtex4-1}

\usepackage{graphicx}% Include figure files
\usepackage{dcolumn}% Align table columns on decimal point
\usepackage{bm}% bold math
\begin{document}

%\preprint{APS/123-QED}

\title{Upper Bound on Fidelity of Classical Sagnac Gyroscope}

\author{Thomas B. Bahder}
\affiliation{%
Aviation and Missile Research, 
      Development, and Engineering Center, \\    
US Army RDECOM, 
Redstone Arsenal, AL 35898, 
U.S.A.}%

\date{\today}

\begin{abstract}
Numerous quantum mechanical schemes have been proposed that are intended to improve the sensitivity to rotation provided by the classical Sagnac effect in gyroscopes.  A general metric is needed that can compare the performance of the new quantum systems with the classical systems. The fidelity (Shannon mutual information between the measurement and the rotation rate) is proposed as a metric that is capable of this comparison.   A theoretical upper bound is derived for the fidelity of an ideal classical Sagnac gyroscope.  This upper bound for the classical Sagnac gyroscope should be used as a benchmark to compare the performance of proposed enhanced classical and quantum  rotation sensors. In fact, the fidelity is general enough to compare the quality of two different apparatuses (two different experiments) that attempt to measure the same quantity.      
\end{abstract}

%Use showkeys class option if keyword

\maketitle

%\tableofcontents

%\section{\label{Intro}Introduction}

The Sagnac effect~\mbox{\cite{Sagnac1913a,Sagnac1913b,Sagnac1914,Post1967}} is the basis for all modern rotation sensors~\cite{Lefevre1983} and their applications to inertial navigation systems~\cite{Titterton2004}. Besides its practical applications, the Sagnac effect is being contemplated for studying general relativistic effects, such as Lense-Thirring frame dragging, detecting gravitational waves, and testing local Lorentz invariance~\cite{Chow1985,Stedman1997}.  

\begin{figure}[t]   
\includegraphics[width=3.0in]{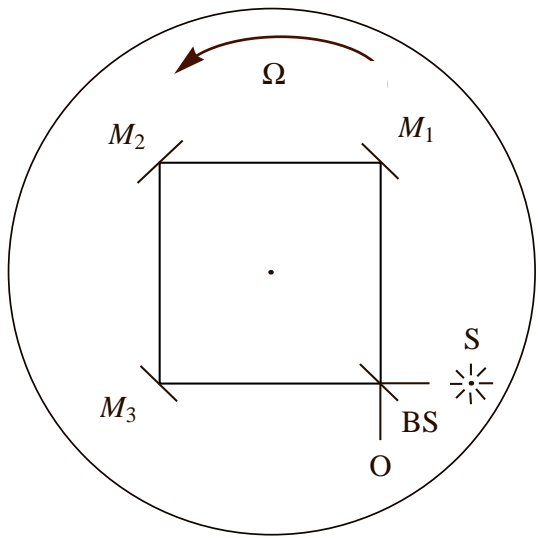}
\caption{\label{fig:SagnacInterferometer}Schematic of a Sagnac interferometer is shown, with light from source $S$ incident on beam splitter $BS$,  mirrors at $M_1$, $M_2$, and $M_3$, and the observer at $O$ that detects the frequency shift $\Delta \omega$.}
\end{figure}

The original experiments of Sagnac consisted of mirrors mounted on a rotating disk, see Fig.~\ref{fig:SagnacInterferometer}.  The mirrors define two paths, one clockwise (CW) and the other counter-clockwise (CCW) on the disk.  A source of light  at point $S$ was mounted on the rotating disk, having wavelength $\lambda = 2 \pi c /\omega$, where $\omega$ is the angular frequency as measured in an inertial frame and $c$ is the speed of light in vacuum. The beam is split at the beam splitter $BS$ and light is propagated along the clockwise and counter-clockwise paths. The two beams are brought together at the beam splitter $BS$ and observed at point $O$.  When the interferometer is rotated at angular velocity $\Omega$, with light source and detector mounted on the rotating disk, a fringe shift  $\Delta N$ is observed with respect to the fringe position for the stationary interferometer, given by~\cite{Post1967,Chow1985}
\begin{equation}
\Delta N  = \frac{4 {\bf A}\cdot {\bf \Omega} }{ \lambda \, c} 
\label{SagnacEffectFringeShift}
\end{equation}
where  ${\bf A}$ is a vector perpendicular to the area enclosed by the paths, having magnitude $A=|{\bf A} |$, and   ${\bf \Omega}$ is the vector in  the direction of the angular velocity of rotation, with magnitude $| {\bf \Omega} | = \Omega$. 
The fringe shift, $\Delta N =\Delta L / \lambda = c \Delta t / \lambda$, can be expressed in terms of the path length difference, $\Delta L$, or  time difference, $\Delta t = 4 {\bf A}\cdot {\bf \Omega}  /  c^2 $, for the CW and CCW paths, as measured in an inertial frame. For typical rotation rates in the laboratory, the classical Sagnac effect is small, and the effect  has to be enhanced to make a practical rotation sensor.

The classical Sagnac effect is exploited for sensing rotation by either measuring a frequency shift or a phase shift. In the active ring laser gyroscope (RLG), where the optical medium is inside the cavity~\cite{Chow1985}, or in a resonant fiber-optic gyroscope (R-FOG)~\cite{Lefevre1983}, a measurement is made of the frequency shift, $\Delta \omega$, between the CW and CCW propagating modes~\cite{Post1967,Chow1985} 
\begin{equation}
\Delta \omega  = \frac{4 A  \,\Omega }{ L  \, c} \omega = S \, \Omega
\label{FrequencyShiftSagnacLaserGyro}
\end{equation}
where $L$ is the length of the perimeter of the path measured in an inertial frame, and $S$ is commonly called the scale factor.  

In a passive fiber ring interferometer (I-FOG)~\cite{Lefevre1983}, or a passive ring laser gyroscope with light source outside the medium~\cite{Chow1985}, the phase shift $\Delta \phi$, is measured between CW and CCW propagating beams,
\begin{equation}
\Delta \phi  = \frac{4 A  \Omega }{ c^2} \omega
\label{SagnacPhaseShift}
\end{equation}
For a fiber-optic gyroscope with phase shift enhanced by $N$ turns, the frequency shift is $\Delta \phi_N = N \Delta \phi$.

Recently, much effort has been expended on experiments with quantum Sagnac interferometers, using single-photons~\cite{Bertocchi2006}, using cold atoms~\cite{Gustavson2000,Gilowski2009} and using Bose-Einstein condensates(BEC)~\cite{Gupta2005,Wang2005,Tolstikhin2005}, in efforts to make improvements over the sensitivity to rotation of the classical Sagnac effect.  Schemes have also been proposed to improve the sensitivity of rotation sensing using multi-photon correlations~\cite{Kolkiran2007} and using entangled particles, which are expected to have Heisenberg limited precision that scales as $1/N$, where $N$ is the number of particles~\cite{Cooper2010}.  

The utility of these quantum systems as rotation sensors must be compared with the classical Sagnac effect using classical light.  The metric used to compare the classical and quantum systems must be sufficiently general to treat both types of systems on an equal basis.  Information measures are examples of such metrics because they are general enough to compare quantum and classical systems.

The determination of the rotation rate is a specific example of the more general problem of parameter estimation, whose goal is to determine one or more parameters from measurements~\cite{Cramer1958,Helstrom1967,Helstrom1976,Holevo1982,Braunstein1994,Braunstein1996,Barndorff-Nielsen2000,Barndorff-Nielsen2003}.  The Cram\'er-Rao theorem\cite{Cramer1958,Cover2006} can be applied to get an upper bound on the variance of an unbiased estimator of a parameter of interest (here the rotation rate) in terms of the classical and quantum Fisher information, see Ref.~\cite{Bahder2010a} and references contained therein. One potential drawback of this approach is that the Fisher information, and therefore the upper bound on the variance of the estimator, can depend on the true value of the parameter to be determined~\cite{Bahder2010a}.  Of course, the true value of the parameter is unknown.  Specifically, the Fisher information can depend on the true angular rotation rate when the Sagnac interferometer is not unitary, which occurs when scattering or dissipation are present~\cite{Bahder2010a}.   Consequently, I propose to characterize a rotation sensor by its fidelity, which is the Shannon mutual information, a quantity that does not depend on the true rotation rate.

In this letter, I calculate a fundamental upper bound on the rotation sensitivity of a classical Sagnac gyroscope that follows Eq.~(\ref{FrequencyShiftSagnacLaserGyro}).  This upper bound can be used as a benchmark to compare the performance of rotation sensors based on newly proposed quantum and classical methodologies, see earlier discussion.  In addition, the fidelity is a useful measure to compare other proposed gyroscopes based on slow light generated by electromagnetically induced transparency~\cite{PhysRevA.62.055801} and other classical optical enhancements~\cite{Matsko2004,PhysRevA.78.053824,PhysRevA.80.011809}.

The fidelity~\cite{Bahder2006,Bahder2010a} is the Shannon mutual information~\cite{Shannon1948,Cover2006} between the measurement (the frequency shift), $\Delta \omega$,  and the parameter to be measured (the rotation rate) $\Omega$: 
\begin{eqnarray}
H & = & \int_{-\infty}^{+\infty} d({\Delta \omega}) \,\, \int_{-\infty}^{+\infty}d \Omega\,\, p(\Delta \omega |\Omega ) \,  p(\Omega) \,\, \times \,    \nonumber  \\
  &  & \log_{2}\left[  \frac{ p(\Delta \omega |\Omega)\,}{\int_{-\infty}^{+\infty}%
\,\,\,p( \Delta \omega|\Omega^{\prime} )  \, p(\Omega^\prime) \,\,d\,\Omega^{\prime}}\right].
\label{ShannonMutualInformation}%
\end{eqnarray}
where $p(\Delta \omega |\Omega )$ is the conditional probability density of measuring  a frequency shift $\Delta \omega$, given a true rotation rate $\Omega$.   Our prior information on the rotation rate is given by the probability density $p(\Omega)$.   The fidelity,  $H$,  is the Shannon mutual information in a communication problem between Alice and Bob, wherein Alice sends messages to Bob~\cite{Shannon1948,Cover2006}.    The fidelity $H$ does not depend on the measurements, $\Delta \omega$, or on the parameter, $\Omega$,  because it is an average over all possible measurements and parameter values.   If we are completely ignorant of the rotation rate, we can take the prior probability as flat distribution, $p(\Omega)=1/(2 \pi)$, indicative of no prior information on our part.  In this case, the fidelity $H$ characterizes the quality of the Sagnac interferometer itself, in terms of mutual information that the measurement of $\Delta \omega$ carries about the parameter of interest, the rotation rate $\Omega$.  In fact, the fidelity $H$ is a specific example of a general way to characterize the quality of all physical measurements.

The fidelity in Eq.(\ref{ShannonMutualInformation}) is a completely general way to describe any classical or quantum measurement experiment. The  classical or quantum apparatus is viewed as a channel through which information flows from the phenomena to be measured to the measurements.  The fundamental quantity that describes this process is the conditional probability of a measurement and the probability distribution that describes our prior information, above notated as $p(\Delta \omega |\Omega )$ and $p(\Omega^\prime)$, respectively.  In the language of communication, there is mutual information $H$ between the continuous alphabet of the parameter, $\Omega$, and the continuous alphabet of the measurements, $\Delta \omega$.

In order to compute the fidelity for the classical Sagnac gyroscope from Equation~(\ref{ShannonMutualInformation}) a model is needed for the conditional probability density $p(\Delta \omega |\Omega )$. In the case of a quantum system, these probabilities are given by a trace:  
\begin{equation}
p(\Delta \omega |\Omega, \rho ) = \rm{tr} \left( \hat{\rho} \,\hat{\Pi}  \left( \Delta \omega \right) \right) 
\label{MeasurementProbabilityDensityMatrix}
\end{equation}
where the state is specified  by the density matrix, $\hat{\rho}$, and the measurements are defined by the positive-operator valued  measure (POVM),  $\hat \Pi( \Delta \omega)$. 

For a classical Sagnac system, I obtain an upper bound on the fidelity in Eq.(\ref{ShannonMutualInformation}).  I consider classical light of bandwidth $\Delta \omega$ and center frequency $\bar{\omega}$,  input into a Sagnac gyroscope that satisfies Eq.(\ref{FrequencyShiftSagnacLaserGyro}).  Therefore, I define the classical measurement probabilities, $p(\Delta \omega |\Omega )$,  by
\begin{equation}
p(\Delta \omega |\Omega ) = \sum\limits_{n = 0}^\infty  {p\left( {\Delta \omega |\Omega ,\omega_n  } \right)\,P_{in} \left( {\omega _n } \right)} 
\label{ClassicalMeasurementProbability}
\end{equation}
where $p\left( {\Delta \omega |\Omega ,\omega_n  } \right)$ is the probability density for measuring $\Delta \omega$, given the that the true rotation rate is $\Omega$, and the input was monochromatic at frequency $\omega_n$. In Eq.~(\ref{ClassicalMeasurementProbability}), for convenience, I assume that the allowed frequency modes are discrete, $\omega_n$, for $n=0,1,\cdots \infty$.   The probability $P_{in} \left( \omega  \right)$ gives the distribution of input frequencies, which has center frequency $\bar{\omega}$ and bandwidth $\Delta \omega$.  As an example, I can take $P_{in} \left( \omega  \right)$ to be a Gaussian distribution of input  frequencies with mean $\bar{\omega}$ and standard deviation $\sigma_\omega$
\begin{equation}
P_{in} \left( {\omega _n } \right) = \left( {\frac{{\delta \omega }}{{2\pi \bar \omega }}} \right)^{1/2} \,\exp \left[ { - \frac{{\left( {\omega _n  - \bar \omega } \right)^2 }}{{2\,\delta \omega \,\bar \omega }}} \right]
\label{GaussianDistribution}
\end{equation}
where $\delta \omega  = \omega _{n + 1}  - \omega _n $ and the variance is given by $\sigma _\omega ^2  = \delta \omega \,\bar \omega $. The distribution of frequencies, $P_{in} \left( {\omega _n } \right)$, is normalized
\begin{equation}
\sum\limits_{n = 0}^\infty  {P_{in} \left( {\omega _n } \right) = 1}
\label{Normalization}
\end{equation}
in the limit $\bar \omega /\delta \omega  \gg 1$.  The size of bandwidth, $\sigma _\omega$, is due to fundamental physical processes in the experiment. 

I want to obtain an upper bound on the fidelity in Eq.(\ref{ShannonMutualInformation}) for a classical system.  Therefore, I assume that classical measurements are have no noise and no bias.  The classical measurement probability, $p(\Delta \omega |\Omega )$,  in Eq.~(\ref{ClassicalMeasurementProbability}) is obtained from Eq.~(\ref{FrequencyShiftSagnacLaserGyro}) as
\begin{equation}
p\left( {\Delta \omega |\Omega ,\omega } \right) = \delta \left( {\Delta \omega  - \frac{{4A\omega }}{{Lc}}\Omega } \right)
\label{ClassicalMeasurementProb}
\end{equation}
where $\delta(x)$ is the Dirac delta function. Using Eq.~(\ref{ClassicalMeasurementProb}) in Eq.(\ref{ClassicalMeasurementProbability}) gives the classical probability of  measuring  $\Delta \omega$ given the true rotation rate is $\Omega$:
\begin{equation}
p(\Delta \omega |\Omega ) = \left| {\frac{{Lc}}{{4A\Omega }}} \right|\,P_{in} \left( {\frac{{Lc}}{{4A{\kern 1pt} \Omega }}\,\Delta \omega } \right)
\label{ClassicalMeasurementFreqProb}
\end{equation}
Note that Eq.~(\ref{ClassicalMeasurementFreqProb}) is valid for an arbitrary input frequency distribution $P_{in} \left( {\omega  } \right)$.  As an example, for a monochromatic frequency 
$\bar{\omega}$ input, 
\begin{equation}
P_{in} \left( \omega  \right) = \delta \left( {\omega  - \bar \omega } \right)  
\label{MonochromaticInput}
\end{equation}
Eq.(\ref{ClassicalMeasurementFreqProb}) gives the probability of classical measurement as expected:
\begin{equation}
p\left( {\Delta \omega |\Omega } \right) = \delta \left( {\Delta \omega  - \frac{{4A{\kern 1pt} \Omega }}{{Lc}}\bar \omega } \right)
\label{ClassicalMeasurementDirac}
\end{equation}
For classical light input, with the Gaussian distribution in Eq.~(\ref{GaussianDistribution}), Eq.(\ref{ClassicalMeasurementFreqProb}) gives the probability of classical measurement as
\begin{equation}
p\left( {\Delta \omega |\Omega } \right) = \left( {\frac{1}{{2\pi }}} \right)^{1/2} \frac{{Lc}}{{4A\left| \Omega  \right|{\kern 1pt} \sigma _\omega  }}\exp \left[ { - \frac{{\left( {\frac{{Lc}}{{4A{\kern 1pt} \Omega }}\Delta \omega  - \bar \omega } \right)^2 }}{{2{\kern 1pt} \sigma _\omega ^2 }}} \right] 
\label{ClassicalMeasurementGaussianInput}
\end{equation}
The conditional probability density in Eq.~(\ref{ClassicalMeasurementGaussianInput}) can be inverted by using Bayes' rule
\begin{equation}
p\left( {\Omega |\Delta \omega } \right) = \frac{{p\left( {\Delta \omega |\Omega } \right)p\left( \Omega  \right)}}{{\int\limits_{ - \infty }^{ + \infty } {p\left( {\Delta \omega |\Omega '} \right)p\left( {\Omega '} \right)\,d\Omega '} }}
\label{BayesRule}
\end{equation}
where $p\left( \Omega  \right)$ specifies the prior probability distribution on rate of rotation, based on our prior information on the rotation rate.  
With the probability distribution in Eq.~(\ref{ClassicalMeasurementGaussianInput}), the conditional probability distribution $p\left( {\Omega |\Delta \omega } \right)$ 
defined by  Eq.~(\ref{BayesRule}) has a divergence.  However, our prior information on the rotation rate, given by the distribution $p\left( \Omega  \right)$ provides a natural cutoff on the integral in Eq.~(\ref{BayesRule}).  We can be reasonably sure that $p( \Omega ) \rightarrow 0$ as $|\Omega | \rightarrow \pm \infty$.  For example, we can take 
\begin{equation}
p\left( \Omega  \right) = \left\{ {\begin{array}{*{20}c}
   {\frac{1}{{2\Omega _{\max } }},} & { - \Omega _{\max }  <  \Omega  <  + \Omega _{\max } }  \\
   {0,} & {{\rm{otherwise}}}  \\
\end{array}} \right.
\label{Cuttoff}
\end{equation}
where $\Omega_{\max}$ represents the maximum expected rotation rate on physical grounds. 

For the monochromatic input frequency in Eq.~(\ref{MonochromaticInput}), the probability of rotation is
\begin{equation}
p\left( \Omega |\Delta \omega  \right) = \delta \left( \Omega  - \frac{L c}{4A}  \frac{\Delta \omega }{\bar \omega } \right)
\label{monRotationRate}
\end{equation}

For the Gaussian frequency distribution in Eq.~(\ref{GaussianDistribution}), Eq.~(\ref{BayesRule}) gives the probability of rotation as
\begin{equation}
p\left( {\Omega |\Delta \omega } \right) = \frac{{\bar \omega }}{{\sqrt {2\pi } {\kern 1pt} \sigma _\omega  }}\frac{1}{{\left| \Omega  \right|}}\exp \left[ { - \frac{1}{{2\sigma _\omega ^2 }}\left( {\frac{{Lc{\kern 1pt} \Delta \omega }}{{4A}}} \right)^2 \left( {\frac{1}{\Omega } - \frac{{4A\bar \omega }}{{Lc{\kern 1pt} \Delta \omega }}} \right)^2 } \right]
\label{GaussianRotationRate}
\end{equation}
In the limit $\Omega \rightarrow \infty$, the probability distribution for $\Omega$,  defined by  Eq.(\ref{GaussianRotationRate}) approaches the function
\begin{equation}
p\left( {\Omega |\Delta \omega } \right) = \frac{{\bar \omega }}{{\sqrt {2\pi } {\kern 1pt} \sigma _\omega  }}\frac{1}{{\left| \Omega  \right|}}\exp \left[ { - \frac{1}{2}\left( {\frac{{\bar \omega }}{{\sigma _\omega  }}} \right)^2 } \right]
\label{GaussianRotationRateLIMIT}
\end{equation}
and hence it is not a normalizable probability distribution because its integral diverges logarithmically like  $\log \Omega $ for $\Omega \rightarrow \infty$. However, this divergence is multiplied by the exponentially small factor 
\begin{equation}
\frac{{\bar \omega }}{{\sigma _\omega  }}\exp \left[ { - \frac{1}{2}\left( {\frac{{\bar \omega }}{{\sigma _\omega  }}} \right)^2 } \right] \ll 1  
\label{ExponentiallSmallFactor}
\end{equation}
since $\bar \omega /\sigma _\omega   \gg 1$.  Note that the peak in the probability distribution for $\Omega$ in Eq.~(\ref{GaussianRotationRate}) occurs at a value $\bar{\Omega}= L c \Delta \omega /(4 A \bar{\omega})$, which is consistent with Eq.(\ref{FrequencyShiftSagnacLaserGyro}).
The probability distribution for the rotation rate in Eq.~(\ref{GaussianRotationRate}) is not a Gaussian distribution.  However it is possible to define a width, $\sigma_\Omega$, that depends on $\Omega$:
\begin{equation}
\sigma _\Omega   = \frac{{\sigma _\omega  }}{{\bar \omega }}\Omega 
\label{RoatationRateWidth}
\end{equation}
Equation~(\ref{RoatationRateWidth}) gives the uncertainty in the rotation rate, $\sigma _\Omega$, in terms of the true rotation rate, $\Omega$, the bandwidth of the input classical light, $\sigma_\omega$, and the center frequency, $\bar{\omega}$, used in the classical Sagnac gyroscope.    As expected, the uncertainty in the rotation rate, $\sigma _\Omega$,  is proportional to the bandwidth of the input light, $\sigma _\omega$.  The uncertainty also decreases with higher input frequency, ${\bar \omega }$. 

The upper bound on the fidelity (Shannon mutual information), $H_{\max}$, for the classical Sagnac gyroscope is given by Eq.~(\ref{ShannonMutualInformation}) using Eq.~(\ref{ClassicalMeasurementGaussianInput}) and Eq.~(\ref{Cuttoff}):
\begin{equation}
H_{\max }  = \frac{1}{2}{\kern 1pt} \log _2 \left[ {\left( {\frac{e}{{2\pi }}} \right)^{1/2} \frac{{\bar \omega }}{{\sigma _\omega  }}} \right]
\label{FidelityMaximum}
\end{equation}
Equation~(\ref{FidelityMaximum}) represents a fundamental theoretical upper bound on the information (in bits) that an ideal classical Sagnac gyroscope can provide to a user, for each measurement of frequency shift, $\Delta\omega$.  The value in Eq.~(\ref{FidelityMaximum}) is an upper bound because we have assumed an ideal classical measurement that has no associated noise.  Therefore, the upper bound in Eq.~(\ref{FidelityMaximum}) for the  classical Sagnac gyroscope is a benchmark to which we can compare rotation sensors based on new quantum technologies, see references above.

In summary, I have proposed the use of a new metric, the Shannon mutual information (called the fidelity) between the rotation rate and the measurements (frequency shift) to judge the quality of a rotation sensor. The fidelity metric is general enough to allow comparison of classical and quantum rotation sensors.  For an ideal classical Sagnac gyroscope,  I have computed a theoretical upper bound on the mutual information that the gyroscope can give to a user by assuming a classical measurement model with no noise.  Consequently, $H_{\max }$ in Eq.~(\ref{FidelityMaximum}) is the Shannon capacity of a classical Sagnac gyroscope.   This upper bound is a benchmark to compare the performance of new  rotation sensors based on improved classical and quantum technologies.  In addition, in Eq.~(\ref{RoatationRateWidth}) I have derived a relation between the bandwidth of light input into a classical Sagnac gyroscope, $\sigma_\omega$, and an estimate of the uncertainty in the rotation rate, $\sigma_\Omega$.  

Finally, the fidelity defined in Eq.~(\ref{ShannonMutualInformation}) is general enough to describe the quality of any physical measurement.  Consequently, the fidelity can be used to compare the quality of two different apparatuses (two different experiments) that attempt to measure the same quantity.   

\vspace{-0.10in}

%  \begin{acknowledgments}
% xxxx
%\end{acknowledgments}

%
%%%%%%%%%%%%%%% REFERENCES  %%%%%%%%%%%%%%%%%%%%
\bibliographystyle{apsrev}
\bibliography{References-Quantum}
%%%%%%%%%%%%%%%%%%%%%%%%%%%%%%%%%%%%%%%%%%%%%%%%
%
\end{document}